\documentclass[preprint,showpacs,showkeys,preprintnumbers,amsmath,amssymb]{revtex4}

\usepackage{graphicx}% Include figure files
\usepackage{dcolumn}% Align table columns on decimal point
\usepackage{bm}% bold math

\begin{document}

\preprint{}

\title{Spatial Random Field Models  
Inspired from Statistical Physics \\ 
with Applications in the Geosciences 
}% Force line breaks with \\
\author{Dionissios T. Hristopulos}
\email{dionisi@mred.tuc.gr } 
 \homepage{http://www.mred.tuc.gr/dionisi.htm}
\affiliation{Department of Mineral Resources Engineering\\
Technical University of Crete\\Chania 73100, Greece}%
\thanks{This work is partially supported by the
EPEAEK Program: Environment Pithagoras II.}
\begin{abstract}
The spatial structure of fluctuations in spatially 
inhomogeneous processes can be modeled in terms of 
 Gibbs random fields.
A local low energy estimator (LLEE) is proposed for the 
interpolation (prediction)
of such  processes
at points where observations are not available.
The LLEE
approximates the spatial dependence of the data and the
unknown values at the estimation points  
by low-lying excitations of a suitable energy functional. 
It is shown that the LLEE is a linear, unbiased, non-exact estimator.
In addition, an expression for the uncertainty 
(standard deviation) of the estimate is derived.
\end{abstract}
\maketitle

\section{Introduction}

Spatial random fields (SRF's) have applications in 
hydrology \cite{kitan,rubin}, oil reservoir engineering
\cite{hohn}, 
environmental pollutant mapping
and risk assessment \cite{christ}, mining exploration and reserves estimation
\cite{goov}, as well as 
environmental health studies \cite{ch98}. 
SRF's model spatial  
correlations in variables such as mineral concentrations, 
dispersion of environmental pollutants, 
soil and rock permeability, 
and flow fields in oil reservoirs. 
Knowledge of spatial correlations enables (i) generating 
predictive iso-level 
contour maps  (ii) estimating the uncertainty of predictions and 
(iii) developing simulations 
that partially reconstruct the process of
interest. Geostatistics provides mathematical tools for these tasks.
The classical approach is based on Gaussian SRF's (GSRF's) and
various generalizations for non-Gaussian distributions
\cite{lantu,wack}. For GSRF's the 
spatial structure is determined from 
the covariance matrix, which is estimated from the distribution 
of the data in space. 

An SRF state (realization) can 
be decomposed into a {\it deterministic trend} $m_{\rm x} ({\bf s})$,  
a {\it correlated fluctuation}  ${X}_{\lambda}({\bf s})$, and an 
independent random noise term, $ \epsilon({\bf s}) $ so that 
$ X({\bf s})=m_{\rm x} ({\bf s})+{X}_{\lambda}({\bf s})+\epsilon({\bf s}).$
The trend represents large-scale variations of the field, 
which can be obtained in principle by ensemble averaging, i.e. 
$m_{\rm x} ({\bf s})=E[X({\bf s})]$.
In practice, the trend is often determined from a single
available realization. 
The fluctuation term corresponds to `fast variations' that reveal 
structure at small scales, which nonetheless exceed a cut-off $\lambda$. 
The random noise represents non-resolved inherent variability 
due to resolution limits, purely 
random additive noise, or  
non-systematic measurement errors. 
It is typically assumed that the 
fluctuation is a {\it second-order stationary SRF}, 
or an {\it intrinsic SRF} with second-order stationary increments
\cite{yaglom}. 
The  {\it observed SRF} after detrending is a
zero-mean fluctuation: 
$ X^{*}({\bf s})=X_{\lambda}({\bf s})+\epsilon({\bf s}).$

In statistical physics the probability density function (pdf) of 
a fluctuation field $x({\bf s})$ governed by an energy 
functional $H[x({\bf s})]$  is  
expressed as
$f_{\rm x} [x({\bf s})] = Z^{- 1} \exp 
\left\{ { - H[x ({\bf s})]} \right\},$
where $ Z $ is the partition function.  
Using this representation,  the Gaussian joint pdf 
in classical geostatistics  
is expressed in terms of the functional:

\begin{equation}
\label{covenergy}
H[x({\bf s})] = \frac{1}{2} \int {d{\bf s}} \int {d{\bf s'}} 
x ({\bf s})\, [G_{x}]^{ - 1} ({\bf s}-{\bf s'})\,
 x  ({\bf s'}).
\end{equation}

\noindent 
In Eq.~(\ref{covenergy}), $[G_{x}]^{ - 1} ({\bf s}-{\bf s'})$ 
is the inverse of the 
 covariance function 
$ G_{\rm x} ({\bf s}-{\bf s'}) $,
which determines the 
{\it spatial disorder}.
While statistical physics plays an increasingly important role
in understanding the behavior of complex geophysical systems \cite{sornette}, 
its applications in geostatistical analysis have not yet been explored.

Spartan Spatial Random Fields (SSRF's) 
model spatial correlations 
in terms of `interactions', in the spirit of 
Markov SRF's \cite{winkler}.
In \cite{dth03} general properties and permissibility conditions were derived
for the fluctuation-gradient-curvature (FGC) SSRF model,
with the following energy functional:

\begin{equation}
\label{fgc}
H_{\rm fgc} [X_\lambda  ] = \frac{1}{{2\eta _0 \xi ^d }}\int {d{\bf s}} \,
\left\{
\left[ {X_\lambda ({\bf s})} \right]^2  + \eta _1 \,\xi ^2 
\left[ {\nabla X_\lambda ({\bf s})} \right]^2  + \xi ^4 
\left[ {\nabla ^2 X_\lambda ({\bf s})} \right]^2 
\right\}.
\end{equation}

For this model,
a moment-based method for parameter estimation
was proposed and tested with simulated data; 
methods for SSRF non-constrained simulation  
were presented in \cite{dth03b};
systematic reduction of anisotropic disorder,
based on the covariance tensor identity, 
was investigated in \cite{dth02,dth04}.
The FGC model 
\cite{dth03} has three main parameters: 
the scale coefficient $\eta_0$, the covariance-shape coefficient
$\eta_1$, and the correlation length $\xi$.  
{\it Bochner's theorem} \cite{christ} for the covariance function 
requires $\eta _1  >  -2$.
A coarse-graining kernel is used to cut off the fluctuations 
at $k_c  \propto \lambda ^{-1} $ 
 \cite{dth03,dth03b}, leading to band-limited
covariance spectral density and differentiable field configurations 
(in the mean square sense) \cite{dth03b}.  

\section{Operator Notation}
\label{hami-not}

Let $\Omega \in {\mathbb R}^{d}$ denote the area of interest and $A(\Omega)$ its
boundary. Consider an SSRF defined over this area with parameters 
$\eta_{0}, \eta_{1}, \xi$, with a finite variance $\sigma^{2}_{\rm x}$.
Let us assume that it is possible to normalize the SSRF to unit variance by simply
dividing the states with the standard deviation.
Next, it is possible to express the pseudo-energy functional 
in terms of an operator notation 
notation as follows: 

\begin{equation}
\label{Hop}
H [X_\lambda  ] \equiv \langle X_\lambda ({\bf s}) |
\, {\cal H} \, | X_\lambda ({\bf s}) \, \rangle +S(A)
\equiv \int_{\Omega} d{\bf s} \, X_\lambda ({\bf s}) \,
{\cal H} \left[ X_\lambda ({\bf s}) \right] + S(A),
\end{equation}

\noindent where ${\cal H}$ is a `pseudo-hamiltonian' operator
and $S(A)$ is a surface term.  
Assuming that the surface term is negligible, the eigenvalue equation 
becomes: 

\begin{equation}
\label{eigv1}
 {\cal H} \, | \psi_{E}({\bf s};{\bf b}) \rangle
=E \,  \psi_{E}({\bf s};{\bf b}) ,
\end{equation}

\noindent
where $\psi_{E}({\bf s};{\bf b})$ is an eigenfunction, 
$E$ is the corresponding energy and ${\bf b}$ a degeneracy vector index,
which may include both discrete and continuous components. 
Since the SSRF has been normalized to unit variance,
the eigenfunctions $\psi_{E} ({\bf s};{\bf b})$ can also be assumed
 normalized, i.e., 
$\int_{\Omega} d{\bf s} \, \psi_{E}^{2}({\bf s};{\bf b})=1$, and then
$H [X_\lambda  ]= E$.

If Eq.~(\ref{eigv1}) admits solutions for non-zero $E$, one can construct 
eigenfunctions that correspond to positive excitation energies $E$. 
The realization probability  that corresponds to low-lying 
excitations is high. 
Hence, the main idea is to 
consider the observed state 
or the union of the observations and the predictions 
as being locally  
represented by an excited state. 
This approach can be used for both parameter estimation and 
prediction (spatial estimation)

\subsection{Eigenfunctions for FGC case} 
\label{fgc_sol}
For the FGC functional of Eq.~(\ref{fgc}), 
integrating the square-gradient term by parts leads to the following equation: 

\begin{equation}
\label{h1}
\int_{\Omega} d{\bf s}\,  
\left[ \nabla \psi_{E}({\bf s};{\bf b})\right]^2 = 
-\int_{\Omega} d{\bf s}\, \psi_{E}({\bf s};{\bf b}) \,
\nabla^2 \psi_{E}({\bf s};{\bf b})
+ \int_{A(\Omega)} d{\bf a} \cdot \nabla \psi_{E}({\bf s};{\bf b}) \, 
\psi_{E}({\bf s};{\bf b}).
\end{equation}

\noindent
In Eq.~(\ref{h1}) $\int _{A(\Omega)} d{\bf a}$ denotes the surface integral 
on the boundary of the area of interest.
Secondly, using Green's theorem on the square-curvature term one obtains 
\begin{eqnarray}
\label{h2}
\int_{\Omega} d{\bf s}\, \left[ \nabla^{2} \psi_{E}({\bf s};{\bf b}) \right]^2 
& = & \int_{\Omega} d{\bf s}\, 
\psi_{E}({\bf s};{\bf b}) \nabla^{4} \psi_{E}({\bf s};{\bf b})
+ \int_{A(\Omega)} d{\bf a}\, \cdot \nabla \psi_{E}({\bf s};{\bf b}) 
\nabla^{2} \psi_{E}({\bf s};{\bf b})
\nonumber \\
& - & \int_{A(\Omega)} d{\bf a}\, \cdot \nabla \left[\nabla^{2} 
\psi_{E}({\bf s};{\bf b}) \right] \psi_{E}({\bf s};{\bf b}) .
\end{eqnarray}

\noindent
Hence, in the operator notation the FGC functional 
is expressed as follows:

\begin{equation}
\label{Hop2}
{\cal H}_{\rm fgc}= \frac{1}{2\eta _0 \xi ^d }
\left[ 1 -\eta_{1} \, \xi^2 \, \nabla^2 + \xi^4  \, \nabla^{4} \right],
\end{equation}

\noindent
and the surface term is given by: 

\begin{eqnarray}
\label{surface}
S(\Omega)& = & \frac{ 1}{2\, \eta_{0}\, \xi^{d}} 
\left[ \eta_{1} \, \xi^2 \,
\int_{A(\Omega)} d{\bf a} \cdot \nabla \psi_{E}({\bf s};{\bf b}) 
\, \psi_{E}({\bf s};{\bf b}) \right.
\nonumber \\
& + & \xi^4 \, \int_{A(\Omega)} d{\bf a}\, \cdot \nabla \psi_{E}({\bf s};{\bf b}) 
\nabla^{2} \psi_{E}({\bf s};{\bf b})
\nonumber \\
         & -  & \left. \xi^4 \, \int_{A(\Omega)} d{\bf a}\, 
       \cdot \nabla \left[\nabla^{2} 
\psi_{E}({\bf s};{\bf b}) \right] \psi_{E}({\bf s};{\bf b}) \right].
\end{eqnarray}

\noindent
If the units are chosen so that 
$2\eta _0 \xi ^d=1$ and the surface term is ignored, 
the eigenvalue equation is given by the following 
partial differential equation (pde): 

\begin{equation}
\label{eigv2}
 \psi_{E}({\bf s};{\bf b}) -\eta_{1} \, \xi^2 \, \nabla^2 \, 
 \psi_{E}({\bf s};{\bf b})
 + \xi^4  \, \nabla^{4} \, \psi_{E}({\bf s};{\bf b}) 
=E \,  \psi_{E}({\bf s};{\bf b}) .
\end{equation}

\noindent
The eigenfunctions $\psi_{E}({\bf s};{\bf b})$ of Eq.~(\ref{eigv2}) are given by
the following four plane waves:
\begin{equation}
\label{eigens}
\psi_{E}({\bf s};{\bf b})= \,e^{{\bf k}_j \cdot {\bf s}},
\quad  {\bf k}_j =k_j \, {\hat{\bm \theta}}, 
\end{equation}

\noindent
where ${\hat{\bm \theta}}$ represents the unit direction vector, 
and $k_j $ the magnitudes of the \em{characteristic wave-vectors} 
that are given by the roots of the fourth-order \em{characteristic polynomial}: 
\begin{equation}
\label{wavvec:char}
\Pi_{\rm fgc} (k \xi) = (1-E)\, - \,\eta _1 \,\xi ^2\,k^2\,+\xi ^4\,\,k^4=0.
\end{equation}

\noindent
Thus, the characteristic wavevectors are given by  
the following expressions:
\begin{eqnarray}
\label{k1}
k_1(\eta_{1},\xi,E) & = & \frac{1}{\sqrt{2}\xi }\sqrt {\eta _1 + \sqrt {\eta _1^2 -4 (1-E)} } 
 \\ 
\label{k2}
 k_2(\eta_{1},\xi,E)& = & - \frac{1}{\sqrt{2}\xi }\sqrt {\eta _1 + \sqrt {\eta _1^2 -4 (1-E)} } 
 \\
 \label{k3}
 k_3(\eta_{1},\xi,E) & = &  \frac{1}{\sqrt{2}\xi }\sqrt {\eta _1 - \sqrt {\eta _1^2 -4 (1-E)} } 
 \\
 \label{k4}
 k_4(\eta_{1},\xi,E) & = & - \frac{1}{\sqrt{2}\xi }\sqrt {\eta _1 - \sqrt {\eta _1^2 -4 (1 - E)} }. 
\end{eqnarray}

\noindent
Note that only the magnitude of the wave-vectors is determined from the 
pde~(\ref{eigv2}). This is due to the fact that isotropic spatial
dependence was assumed in the SSRF model. 

(a) If $\eta_{1} >0 \wedge 1-\eta_{1}^2/4 <E <1$ all the roots are real. 
(b) If $\eta_{1} >0 \wedge E>1$, then $k_{1}, k_{2}$ are real,
while $k_{3}, k_{4}$ are purely imaginary.
(c) If $\eta_{1} >0 \wedge 1-\eta_{1}^2/4 >E$, then all the roots are complex. 
(d) If $\eta_{1} <0 \wedge 1-\eta_{1}^2/4 <E <1$, then 
all the roots are imaginary. 
(e) If $\eta_{1} <0 \wedge E > 1$, then $k_{1}, k_{2}$ are real,
while $k_{3}, k_{4}$ are imaginary.
(f) If $\eta_{1} <0 \wedge 0< E < 1-\eta_{1}^2/4$ all the roots are complex. 

In general, an excited state formed by the linear superposition 
of degenerate eigenstates of energy $E$ is given by the expression:

\begin{equation}
\label{state_E}
Z_{E}({\bf s}; c_{\bf b}) = \sum_{j=1}^{4} 
u\left( k_{c}- \| k_{j}\| \right) \,
\int d\hat{\bm \theta} \,
 c_{j}(\hat{\bm \theta}) \,
\exp \left( k_{j} \, {\hat{\bm \theta}} \cdot {\bf s} \right),
\end{equation}

\noindent
where $c_{j}(\hat{\bm \theta})$ is a direction-dependent
 (possibly complex-valued) function, 
 $\| k_{j} \| $ is the modulus of the characteristic wavevector,
and $u(.)$ is the unit step function, used to guarantee
that the fluctuations in 
the excited state do not exceed the 
cutoff `frequency'.
For the estimation of real-valued processes, 
the coefficients $c_{j}(\hat{\bm \theta})$ are constrained to
 give real values for the excited state $Z_{E}({\bf s}; c_{\bf b})$.
If $c_{j}(\hat{\bm \theta})=c_{j}$, an {\it isotropic excited state} is obtained, 
which can be expressed as  
$Z_{E}({\bf s}; c_{1},\ldots,c_{4})=\sum_{j=1}^{4} c_{j} \, 
u\left( k_{c}- \| k_{j}\| \right)
\psi_{E}({\bf s};j)$, where 
$\psi_{E}({\bf s};j)=\int d\hat{\bm \theta} \exp \left( k_{j} \, {\hat{\bm \theta}} 
\cdot {\bf s} \right)$.

\subsection{Eigenstates in $d=1$}
We examine in more detail the real-valued eigenstates 
that are  trigonometric or hyperbolic functions 
in the one-dimensional domain $[0,\, L] \in {\mathbb R}$. 

\subsubsection{Exponential Eigenstates}
For characteristic wave-vectors $k$ that are real numbers, the 
{\em normalized eigenfunctions} and the corresponding energies 
of Eq.~(\ref{eigv2}) are given by 

\begin{eqnarray}
\label{expo1d}
X(s) & = & e^{-k\,s} \, \sqrt{\frac{2\, k}{1-e^{-2k\,L}}}, 
\\
\label{Eexpo1d}
E & = & 1 - \eta_{1} (k\, \xi)^{2} + (k\, \xi)^{4}.
\end{eqnarray}
\noindent
However, if the exponential function is inserted in Eq.~(\ref{fgc}), 
the resulting energy is given by 
\begin{equation}
\label{Hexpo1d}
H[X(s)]=1 + \eta_{1} (k\, \xi)^{2} + (k\, \xi)^{4}.
\end{equation}
\noindent
The difference between the energy given by 
Eq.~(\ref{expo1d}) and the correct energy,
given by Eq.~(\ref{Hexpo1d}) is due to the fact that the boundary term 
can not be ignored for the localized exponential excitation.  

\subsubsection{Trigonometric Eigenstates}
If $k$ is an imaginary number, the 
eigenfunctions are trigonometric functions. 
A normalized cosine eigenfunction and the corresponding energy are given by: 

\begin{eqnarray}
\label{trig1dX}
X(s) & = & \cos (k\,s) \, \sqrt{\frac{2}{L \, \left[1 +{\rm sinc}(2k\,L) \right]}}, 
\\
\label{trig1dE}
E& = & 1 + \eta_{1} (k\, \xi)^{2}  \,
\frac{1 -{\rm sinc}(2k\,L)}{1 +{\rm sinc}(2k\,L)}
+ (k\, \xi)^{4}.
\end{eqnarray}
\noindent 
For large domains, $k\, L>>1$, Eq.~(\ref{trig1dE}) is practically 
equivalent to Eq.~(\ref{Hexpo1d}).  As expected, in the case of an 
extended eigenstate (as the cosine) the boundary term can be ignored.

\section{Spatial Estimation with SSRF's}
\label{spatest}

Assume $S_{\rm m}=({\bf s}_1, \ldots {\bf s}_N)$ is a 
set of data points with the respective vector of
measurements denoted by 
${\bf X}^{*}=(X^{*}_{1},\ldots,X^{*}_{N})$;
let  ${\bf s}_{0} \notin S_{\rm m}$ be the estimation point  
 and $\hat{X}_{\lambda}({\bf s}_{0})$
the estimate (spatial prediction). 
The local neighborhood  of ${\bf s}_{0}$
is the set $S_{0} \equiv B({\bf s}_{0}; r_c)$
 of all the data points ${\bf s}_{j}, j=1,...,M$ 
 inside a `sphere' of radius equal to  
 one correlation range from ${\bf s}_{0}$. 
In geostatistics,  $\hat{X}({\bf s}_{0})$ is determined by optimal linear 
filters (kriging estimators) \cite{kitan,wack}, which 
form the estimate as a superposition of the data values inside
the local neighborhood, and there is no explicit resolution 
scale. The coefficients of the superposition are 
selected to make the estimate unbiased and to minimize the mean square error. 
Kriging is an exact interpolator, meaning that for any 
${\bf s}_{i} \in S_{\rm m}, \hat{X}({\bf s}_{i}) = X^{*} ({\bf s}_{i})$.
Exactitude is not always desirable, since
it ignores measurement errors and leads to
excessive smoothing of the fluctuations. 
Hence, different estimation methods are useful. 
The SSRF models can be used in kriging algorithms to 
provide new, differentiable covariance functions.
In addition, within the SSRF framework it is possible to 
define a new type of estimator.  

\subsection{Low Local Energy Estimators}
\label{llee}

The central idea is that a `good' estimate should correspond to
 a state with significant probability of realization. 
If the energy functional is non-negative, as in Eq.~(\ref{fgc}),
 the highest probability is associated with the uniform state  
 $X_\lambda({\bf s})=0$, which is not physically interesting.
Other states with high probability correspond to low-energy excitations. 
Let us superimpose the degenerate eigenstates with energy $E$
  to form a {\em mixed state}   
 $Z_{E}({\bf s};{\bf c})= \sum_{i=1}^{D} c_{i}\, \psi_{E}({\bf s};b_{i})$;
 ${\bm c}=(c_{1},\ldots,c_{D}) $ 
 is a $D$-dimensional vector of linear coefficients that 
 correspond to the degeneracy indices. 
 In principle $D$ can be infinite since the 
 directional dependence given by Eq.~(\ref{state_E}) 
 is continuous.  However, in practice it may be simplest to 
 restrict the search to one `optimal' direction.
 The energy $H [Z_{E}({\bf s}; c_{\bf b})]$ 
 of the mixed state is not necessarily equal to $E$.
In fact, for orthonormal eigenstates  
$H [Z_{E}({\bf s}; {\bf c})]= \mu \, E$, where
$\mu= \sum_{i=1}^{D} c_{i}$.  This reflects the fact that the `energy level' of 
the observed process is set by the measurements (i.e., the coefficients 
$c_{i}$). Since the scale coefficient $\eta_{0}$ is inversely proportional to the 
magnitude of the fluctuations, it follows that $\mu^{-1} \propto \eta_{0}$.  
It should also be noted that if two mixed states $({\bf c}_{1},E_{1})$ and  
$({\bf c}_{2},E_{2})$ are energetically equivalent, i.e., 
$\mu_{1} \,E_{1}=\mu_{2} \,E_{2}$, they are not
in general linearly related, since according to 
Eqs.~(\ref{eigens}), (\ref{k1})-(\ref{k4}) and~(\ref{state_E}),
the dependence of $Z_{E}({\bf s};{\bf c})$ on $E$ is nonlinear.
 
 We propose that the observations 
 for ${\bf s}_{j} \in B({\bf s}_{0}; r_c)$  be expressed as 
 $X^{*}({\bf s}_j)=Z_{E}({\bf s}_{j};{\bf c}_{0})
 +\varepsilon({\bf s}_j)$, where
 $Z_{E}({\bf s};{\bf c}_{0})$ is a `local' excitation 
 and $\varepsilon({\bf s}_j)$ is the  {\it local excitation residual}. 
 Local dependence stems from the fact that the 
 coefficients ${\bf c}_{0}$ depend on ${\bf s}_{0}$, in contrast with 
 the solution of Eq.~(\ref{state_E}), in which the coefficient 
 vector is global. 
 The {\em LLEE estimator} is then given by 
 $\hat{X}_{\lambda}({\bf s}_{0})=Z_{E}({\bf s}_{0};{\bf c}_{0})$.
 Since $Z_{E}({\bf s};{\bf c}_{0})$ is an {\em estimate} of the 
 underlying process $X_\lambda({\bf s})$, the excitation residual 
 $\varepsilon({\bf s}_j)$ is not in general the same as the 
 noise $\epsilon ({\bf s})$. 
 The coefficients  ${\bf c}_{0}$, follow from minimizing  
 the mean square excitation residual inside $B({\bf s}_{0}; r_c)$ , i.e.,  
 
 \begin{equation}
 \label{coeff}
{\bf c}_{0}=\underbrace{{\rm arg} \: {\rm min}}_{\bf c} \sum_{j=1}^{M} 
 \left[ X^{*}({\bf s}_j) - Z_{E}({\bf s}_{j};{\bf c}) \right] ^{2}.
 \end{equation}  
 
 \noindent
 The above is a typical problem of multiple linear regression, where the
 regressors are the functions ${\psi}_{E}({\bf s}_{i};b_{j})$.
 If we define the $M \times D$ matrix 
 $\psi_{E,ij} \equiv {\psi}_{E}({\bf s}_{i};b_{j})$,
 the solutions for $c_{0,i}$ and the LLEE are given by:

\begin{eqnarray}
\label{coefsol}
\alpha_{ik} & = & \sum_{j=1}^{M} \psi_{E,ji} \,\psi_{E,jk},
\quad i=1,\ldots, M; \,\, k=1,\ldots, D,
\\
c_{0,i} & = &
\sum_{k=1}^{D} \left[ \alpha \right]^{-1}_{ik} \, \sum_{l=1}^{M} \psi_{E,lk}\, X^{*}_{l}, \quad  i=1,\ldots, M, 
\end{eqnarray}

\begin{equation}
\label{estims0}
\hat X_{\lambda}({\bf s}_{0}) =  {\bf w}_{0} \, \cdot {\bf X}^{*} ,
\end{equation}

\noindent
where  ${\bf w}_{0} $ is a {\it weight vector} given by:
\begin{equation}
\label{weight}
w_{0,i}  =  \sum_{k=1}^{D}\psi_{E,0k} \, 
  \sum_{j=1}^{D} \left[ \alpha \right]^{-1}_{kj} \, \psi_{E,ij}, 
  \quad i=1,\ldots, M.
\end{equation}

The uncertainty of the LLEE estimate is determined from the ensemble
variance of the local excitation residual
$ \sigma_\varepsilon ^2 ({\bf s}_0 ) = 
E\left[ {X^{*} ({\bf s}_0 ) - \hat X ({\bf s}_0 )} \right]^2 $, 
i.e.: 

\begin{equation}
\label{errestim}
 \sigma _\varepsilon ^2 ({\bf s}_0 ) = 
 \sigma_{\rm x^{*}}^2 +
 \sum_{i=1}^{M} \sum_{j=1}^{M} w_{0,i}\, w_{0,j} \, G_{{\rm x}^{*},ij}  
  - 2 \, \sum_{i=1}^{M} w_{0,i} \, G_{{\rm x}^{*},0i} \quad , 
\end{equation}

\noindent
where $G_{{\rm x}^{*},ij} = E\left[ X^{*}_{i} \, X^{*}_{j} \right]$
is the covariance matrix at the observation points,
$G_{{\rm x}^{*},0i}= E\left[ X^{*}_{i} \, X^{*}_{0} \right]$, 
is the covariance vector of the fluctuations between ${\bf s}_{0}$ and the 
estimation point, 
and $\sigma_{\rm x^{*}}^2= E\left[ X^{*}_{i} \,X^{*}_{i} \right]$
is the variance of the observed process.

\subsection{Properties of the LLEE}
It follows from Eqs.~(\ref{estims0}) and~(\ref{weight}) that the LLEE is linear
 in the fluctuations. Hence, the estimates are unbiased and follow 
the Gaussian law (if the observations are normally distributed). 
Kriging methods are based on 
minimization of the (ensemble) mean square error, 
which is a global optimality criterion. 
In contrast, the LLEE  criterion is local (i.e., minimum 
of the average  squared excitation residual in 
the neighbourhood of the estimation point). 
Another difference with kriging is that low local energy estimates
 do not match exactly the measurements
at observation points. The property of non-exactitude is maintained 
even when the noise can be ignored.
Finally, unlike kriging predictions, the LLEE provides 
multiple estimates, 
since different energy levels lead 
to different excited states. In this respect the LLEE is similar to 
a simulation method. 
However, simulations involve the generation of  
random numbers,
in contrast with the LLEE method.  
It should also be noted that the energy of local excitations is not 
necessarily the energy of the estimated state, because the locality
of the coefficient vector ${\bf c}_{0}$ means that
the operators $\nabla$ and $\nabla^{2}$ 
contribute to the overall energy when they act
on the coefficients of the mixed state
in Eq.~(\ref{fgc}).

\section{Conclusions}
A spatial estimation method for applications in 
the geosciences is presented.  
The method is based on the use of `pseudo-energy' functionals, motivated by
explicit constraints or heuristic physical arguments, to 
capture the spatial heterogeneity of the observed process. 
Estimates of the process at unmeasured points (predictions) are  
based on local interpolating functions that
represent low-energy excitations
of the pseudo-energy.
Multiple estimates of the process can be generated by considering 
local interpolating functions that correspond to different 
excitation energies.

% \appendix

% \section{}
% \label{}

\end{document}